\documentstyle[12pt,twoside]{article}
\begin{document}

\title{\Large\bf A late time acceleration of the universe with two scalar 
fields : many possibilities}

\author{Narayan Banerjee\footnote{email: narayan@juphys.ernet.in}~~~and~~
Sudipta Das \\
Relativity and Cosmology Research Centre,\\
Department of Physics, Jadavpur University,\\ Kolkata - 700 032,\\India.
\date{}
}

\maketitle
\vspace{0.5cm}
{\em PACS Nos.: 98.80 Hw
\par Keywords : Cosmology, Accelerated expansion, Scalar fields}
\vspace{0.5cm}

\pagestyle{myheadings}
\newcommand{\be}{\begin{equation}}
\newcommand{\ee}{\end{equation}}
\newcommand{\bea}{\begin{eqnarray}}
\newcommand{\eea}{\end{eqnarray}}

\begin{abstract}
In the present work, an attempt has been made to explain the recent 
cosmic acceleration of the universe with two mutually interacting 
scalar fields, one being the Brans - Dicke scalar field and the 
other a quintessence scalar field. Conditions have been derived for 
which the quintessence scalar field has an early oscillation and it 
grows during a later time to govern the dynamics of the universe.
\end{abstract}

\section{Introduction}
\par The present cosmic acceleration is now generally believed to be a 
certainty rather than a speculation. The recent data on supernovae of 
type Ia suggested this possibility quite strongly \cite{riess, perl1,
perl2, tonry} and 
the most trusted cosmological observations, namely that on Cosmic 
Microwave Background Radiation \cite{mel, lange, jaffe, netter, halver}, 
appear to be quite 
compatible with an accelerated expansion of the present universe. 
The natural outcome of these observations is indeed a vigorous search 
for the form of matter which can give rise to such an expansion, 
as a normal matter distribution gives rise to an attractive gravity 
leading to a decelerated expansion. This particular form of matter, 
now popularly referred to as ``dark energy", is shown to account for 
as much as 70\%  of the present energy of the universe. 
This is also confirmed by the highly accurate Wilkinson 
Microwave Anisotropy Probe (WMAP) \cite{bridle, ben, hin, kog, sper}. 
A large number 
of possible candidates suitable as a dark energy component have 
appeared in the literature. Excellent reviews on this topic are available 
\cite{sahni1, sahni2, peddy}. Most of the dark energy candidates 
are constructed so as to generate an effective pressure which is sufficiently 
negative driving an accelerated expansion. The alleged acceleration can 
only be a very recent phenomenon and must have set in during the 
late stages of the matter dominated expansion of the universe. 
This requirement is crucial for the successful nucleosynthesis in 
the radiation dominated era as well as for a perfect ambience for 
the formation of structure during the matter dominated era. Fortunately, 
the observational evidences are also strongly in favour of a scenario 
in which the expansion of the universe had been decelerated 
(the deceleration parameter $q > 0$) for high redshifts and 
becomes accelerated ($q < 0$) for low values of the redshift 
$z$ \cite{ag}. So the dark energy sector should have evolved in 
such a way that the consequent negative pressure has begun dictating 
terms only during a recent past.
\par This so easily reminds one about the inflationary universe models
 where an early accelerated expansion was invoked so as to
 wash away the horizon, fine tuning and some other associated 
problems of the standard big bang cosmology. The legendary 
problem, that the inflationary models themselves had, was that of a 
``graceful exit"  -  how the accelerated expansion gives way to 
the more sedate ($q < 0$) expansion so that the universe could 
look like as we see it now. For a very comprehensive review, we 
refer to Coles and Lucchin \cite{coles} or Kolb and Turner \cite{kolb}. 
The ``graceful exit" problem actually stems from the fact that
for the potentials driving inflation, the phase transition to the 
true vacuum  is never complete in a sizeable part of the actual
volume of the universe. The attempts to get out of this problem
involved the introduction of a scalar field which slowly
rolls down its potential so that there is sufficient time available 
for the transition of phase throughout the actual volume of the universe.
 The current accelerated expansion thus poses the problem somewhat
complementary to the graceful exit - that of a ``graceful entry".
\par The present work addresses this problem, and perhaps provides some
clue regarding the solution of the problem. The basic motivation stems 
from existing literature on inflation. Mazenko, Wald and Unruh \cite{wald} 
showed that a classical slow roll is in fact invalid when the 
single scalar field driving inflation is self interacting. 
It was also shown that a slow roll with a single scalar field
puts generic restrictions on the potentials driving inflation \cite{adams}.
 This kind of problems led to the belief that for a successful 
inflationary model, one needs to have two scalar fields \cite{linde}. 
The present work uses this idea of utilising two scalar fields, 
one of them being responsible for the present acceleration of the 
universe and is called the quintessence field. The second one 
interacts nonminimally with the former so that the quintessence 
field has an oscillatory behaviour at the early matter dominated epoch 
but indeed grows later to dominate the dynamics of the more recent 
stages of the evolution. If such a behaviour is achieved, some clue 
towards the resolution of the graceful entry problem or the coincidence 
problem may be obtained. There are quite a few quintessence potentials 
already in the literature \cite{sahni1} which drives a late time 
acceleration, but none of them really has an underlying physics 
explaining their genesis. As one is already hard pressed to find 
a proper physical background of the quintessence field, it will be even 
more embarassing to choose a second field without any physical motivation. 
Naturally, the best arena is provided by a scalar tensor theory, such as 
Brans-Dicke theory, where one scalar field is already there in the 
purview of the theory and is not put in by hand. It deserves mention 
that the Brans-Dicke scalar field was effectively used in 
``extended inflation" in order to get a sufficient slow roll of 
the scalar field \cite{la, johri}. Later Brans-Dicke theory was used 
for finding a solution of the graceful exit problem with a large 
number of potentials \cite{nb}, where the inflaton field evolves 
to an oscillatory phase during later stages.
\par In the next section we write down the field equations in 
Brans-Dicke (B-D) theory with a quintessence field $\phi$, 
the potential $ V(\phi)$ driving acceleration being modulated 
by the B-D scalar field $\psi$ as $ V(\phi)\psi^{-\beta}$.
With a slow roll approximation, the conditions for an initially
 oscillating $\phi$, which grows only during later stages, are 
found out for two examples, a power law expansion and an 
exponential expansion of the scale factor. The particular form of 
$V(\phi)$ is quite irrelevant in this context, the conditions 
only put some restrictions on the constants of the theory and 
the parameters of the model. So the form of the potential is 
arbitrary to start with, only the conditions on the parameters 
and the `value' of $V(\phi)$ has to be satisfied and hence many 
possibilities are opened up to accommodate a physically viable 
potential as the driver of the late time acceleration. However, 
in some cases, this could restrict the form of $V(\phi)$ as well.
In the last section, we make some remarks on the results obtained.
\section{A model with a Graceful Entry}
The relevant action in Brans-Dicke theory is given by
\be
S =\int [\frac{\psi R}{16\pi G_{0}} - 
\omega\frac{{\psi_{,\mu}}{\psi^{,\mu}}}{\psi} - 
\frac{1}{2}{\phi_{,\mu}}{\phi^{,\mu}} - U(\psi, \phi) + 
\it{L}_{m}]\sqrt{-g} d^4x
\ee
where $G_{0}$ is the Newtonian constant of gravitation, $\omega$ is the 
dimensionless Brans-Dicke parameter, $R$ is the Ricci scalar, $\psi$ and 
$\phi$ are the Brans-Dicke scalar field and quintessence scalar fields 
respectively. If we now choose $U(\psi, \phi)$ as $ V(\phi)\psi^{-\beta}$ as 
explained, the field equations, in units where  $8{\pi} G_{0} = 1$ , can 
be written as
\bea
3 H^2 + 3 H \frac{\dot{\psi}}{\psi} - \frac{\omega}{2}\frac{\dot{\psi}^2}{\psi^2} = V(\phi)\psi^{-(\beta + 1)} + \frac{\rho}{\psi}~,\\
 3H \dot{\phi} + V'(\phi)\psi^{-\beta} = 0~,~~~~~~~~~~~~~~~~~~~~~~~~~\\
(2\omega + 3) (\ddot{\psi} + 3H\dot{\psi}) = (\beta - 4) V(\phi) \psi^{-\beta} + \rho~,
\eea
where a dot represents differentiation with respect to time $t$ and a 
prime represents differentiation with respect to the scalar field $\phi$.
As we require the potential $U(\psi, \phi)$ to grow with time so that 
the effective negative pressure dominates at a later stage, $\beta$ should 
be negative for a $\psi$ growing with time or positive for a $\psi$ decaying 
with time. The field equations are written in the slow roll approximation, i.e,
where $\dot{\phi}^2$ and $\ddot{\phi}$ are neglected in comparison to 
others. $ H = \frac{\dot{a}}{a}$ is obviously the Hubble parameter. 
As we dropped the field equation containing stresses, we can use the 
matter conservation equation as the fourth independent equation which 
yields on integration
\be
\rho = \frac{\rho_{0}}{a^3}~,
\ee
$\rho_{0}$ being a constant. This is so as the fluid pressure is taken 
to be zero as we are interested in the matter dominated era.
\par Using the expression for $V(\phi)\psi^{-\beta}$ from equation (2) 
in equation (4), we can write
\be
(2\omega + 3)[\frac{\ddot{\psi}}{\psi} + 3H\frac{\dot{\psi}}{\psi}]
 = (\beta - 4)[ 3H^2 + 3H\frac{\dot{\psi}}{\psi}-\frac{\omega}{2}
\frac{\dot{\psi}^2}{\psi^2}] - ( \beta - 5)\frac{\rho_{0}}{a^3\psi}~.
\ee
If the scale factor $a$ is known, this equation can be integrated 
to yield the Brans-Dicke field $\psi$.\\
Equations (3) and (4) yield
\be
\frac{V(\phi)}{V'(\phi)} = -\frac{(2\omega + 3)(\ddot{\psi} + 3H\dot{\psi}) - \rho}{3(\beta - 4) H\dot{\phi}}~.
\ee
Hence, if we define\\
\be
f(\phi) = \int \frac{V}{V'}d\phi~,
\ee                

then 
\be
f(\phi_{0}) - f(\phi_{i}) = \int_{t_{i}}^{t_{0}} F(t)dt~,                   
\ee
where
\be
F(t) = -\frac{(2\omega + 3) (\ddot{\psi} + 3H\dot{\psi}) - \rho}{3(\beta - 4)H}
\ee
and subscripts `o'  and `i' stands for the present value and some 
initial value, such as the onset of the matter dominated phase of 
evolution.
\par For a given $ a = a(t)$, therefore, equation (6) can be used 
to find $\psi$, which in turn, with equations (8) and (9) determines 
$f(\phi)$. Now, these equations can be used to put bounds on the 
values of derivatives of the potential, which would ensure that 
the dark energy has an oscillating phase in the early stages. 
\par The complete wave equation for the quintessence field $\phi$ is
\begin{center}
$\ddot{\phi} + 3H\dot{\phi} + V'(\phi) \psi^{-\beta} = 0$. \\
\end{center}
The condition for a small oscillation of $\phi$ about a mean value is 
$V'(\phi) = 0$. This provides a kind of plateau for the potential which 
hardly grows with evolution and hence the dynamics of the universe is 
practically governed by the B-D field $\psi$ and the matter density $\rho$. 
We find the condition for such an oscillation of $\phi$ at some initial 
epoch by choosing $V'(\phi) \approx  0$ which yields 
$$
{\frac{\ddot{\phi}}{3H\dot{\phi}}\vline~}_{i} \approx -1~.
$$
During later stages, the universe evolves according to the equations 
(2) - (4) where $V'(\phi) \ne 0$, and the scalar field slowly rolls along 
the potential so that the quintessence field takes an active role in the 
dynamics and gives an accelerated expansion of the universe.
\par Two examples, one for power law and the other exponential expansion, 
will be discussed in the present work.\\
\\
I. Power law expansion :-
\\
\\If $a = a_{0}t^n$ ~where $n > 1$,
 the universe expands with a steady acceleration, i.e, with a constant 
negative deceleration parameter ~$q = -\frac{(n -1)}{n}$.
\\With this, 
\be
H = \frac{\dot{a}}{a} = \frac{n}{t}~,
\ee
and equation (6) has the form
\be
c_{1}\frac{\ddot{\psi}}{\psi} + c_{2}\frac{\dot{\psi}}{\psi}\frac{1}{t} 
+ c_{3}\frac{\dot{\psi}^2}{\psi^2} + c_{4}\frac{1}{t^2} + 
c_{5}\frac{1}{t^{3n}\psi} = 0 ~,
\ee
$c_{i}$'s being constants given by,
\be
c_{1} = (2\omega + 3),~~~
c_{2} = 3n(2\omega - \beta + 7 ),~~~
c_{3} = \frac{\omega}{2}(\beta - 4),~~
c_{4} = -3n^2(\beta - 4),~~ 
c_{5} = (\beta - 5) \frac{\rho_{0}}{a_{0}^3}~.
\ee
The simplest solution for $\psi$ in equation (12) is
\be
\psi = \psi_{0} t^{2 - 3n}~,
\ee
$\psi_{0}$ being a constant.\\
The consistency condition for this is,
$$
(2\omega + 3)(2 - 3n)(1 - 3n) + 3n(2\omega - \beta + 7)(2 - 3n) + 
\frac{\omega}{2}(\beta - 4)(2 - 3n)^2 - 3n^2(\beta -4) + 
(\beta - 5)\frac{\rho_{0}}{a_{0}^3\psi_{0}} = 0~.
$$
From equations (9) and (10), 
\be
f(\phi_{0}) - f(\phi_{i}) = D~(t_{i}^{2 - 3n} - t_{0}^{2 - 3n})~,
\ee
where $D$ is a constant involving $c_{i}$'s, i.e, $n$, $\omega$,
$\psi_{0}$, $\rho_{0}$ etc. given by
$$
D = \frac{(2\omega + 3)\psi_{0}(2 - 3n) - 
\frac{\rho_{0}}{a_{0}^3}}{3n(\beta - 4)(2 - 3n)} ~.
$$
From the form of potential $V = V(\phi)$, $f(\phi)$ can be found 
from the relation (8). So, the different constants will be related 
by equation (15).
\par If the quintessence field oscillates with small amplitude
about the equilibrium at the beginning, i.e, at $t = t_{i}$ and 
grows during the later stages of evolution, then 
${\frac{\ddot{\phi}}{3H\dot{\phi}}\vline~}_{i} \approx 1$, which 
puts more constraints 
amongst different parameters of the theory.\\
Equations (7) and (14) yield
$$
\dot{\phi} = A~t^{(1 - 3n)}~(ln ~V)'~,
$$
where $A$ is a constant given by
$$
A = -\frac{(2\omega + 3)\psi_{0}(2 - 3n) -
\frac{\rho_{0}}{a_{0}^3}}{3n(\beta - 4)} = -D(2 - 3n)~.
$$
So,
$$
\frac{\ddot{\phi}}{3H\dot{\phi}} = -1 + \frac{1}{3n} 
+\frac{A}{3n}~t^{2 - 3n} (ln ~V)'' ~.
$$
The condition for the small oscillation of $\phi$ close to 
$t = t_{i}$ is
$$
{\frac{\ddot{\phi}}{3H\dot{\phi}}\vline~}_{i} \approx -1~,
$$
which gives the condition on $(ln ~V)''$ as
\be
{(ln ~V)''\vline~}_{i} \approx -\frac{1}{A}~t_{i}^{3n - 2}.
\ee
\\
II. Exponential expansion :-
\\
\\Similarly for an exponential expansion at the present epoch, 
the constraints can be derived. For such an expansion,
$$a = a_{0} ~e^{\alpha t}~,$$ 
where $a_{0}$, $\alpha$ are all positive constants.
\\Then,
$$
H = \frac{\dot{a}}{a} = \alpha ~~ and ~~ 
\rho = \frac{\rho_{0}}{a_{0}^3}e^{-3\alpha t}~.
$$
Equation (6) with this solution has the form
\be
b_{1}\frac{\ddot{\psi}}{\psi} + b_{2}\frac{\dot{\psi}}{\psi}
+ b_{3}\frac{\dot{\psi}^2}{\psi^2} + b_{4} +
b_{5}\frac{1}{e^{3\alpha t}\psi} = 0 ~,
\ee
where $b_{i}$'s are constants given by
\be
b_{1} = (2\omega + 3),~~~
b_{2} = 3\alpha(2\omega - \beta + 7 ),~~~
b_{3} = \frac{\omega}{2}(\beta - 4),~~~
b_{4} = -3\alpha^2(\beta - 4),~~
b_{5} = (\beta - 5) \frac{\rho_{0}}{a_{0}^3}~.
\ee
 A simple solution for $\psi$ is
\be
\psi = \psi_{0} e^{-3\alpha t}~,
\ee
$\psi_{0}$ being a constant.
The consistency condition for this is,
\be
9\alpha^2(\beta + \frac{\omega\beta}{2} - 2\omega - 4)
- 3\alpha^2(\beta - 4) + (\beta - 5)\frac{\rho_{0}}{a_{0}^3\psi_{0}} = 0~.
\ee
Equations (9) and (10) will put restrictions on the parameters of 
potential by
\be
f(\phi_{0}) - f(\phi_{i}) = \frac{\rho_{0}}{9\alpha^2a_{0}^3(\beta - 4)}
~[e^{-3\alpha t_{i}} - e^{-3\alpha t_{0}}]~.
\ee
The condition $\frac{\ddot{\phi}}{3H\dot{\phi}} \approx -1$ for the small
oscillation of the scalar field at $t = t_{i}$, with the help of equations 
(7) and (19), 
leads to the interesting result
\be
{(ln ~V)''\vline~}_{i} \approx 0~,
\ee
which indicates that for an exponentially expanding present stage of evolution,
the quintessence potential behaves exponentially at least at the beginning.
\par Thus, if $V = V(\phi)$ is given, then equations (15) and (21) will 
help finding out the bounds on the values of the scalar field $\phi$ and 
the constants appearing in $V(\phi)$ in terms of the initial and final epochs. 
\par The conditions (15) or (21) put bounds on the potential so that the 
quintessence field $\phi$ has an oscillatory behaviour in the beginning. As 
the solutions in the examples considered has accelerated expansion, 
the Q-field $\phi$ has a steady growth at later stages. For a power law 
expansion, the growth has some arbitrariness  as $V(\phi)$ is not specified. 
For the exponential expansion, however, the growth of $\phi$ is governed 
by an exponential potential. 
\section{Conclusion}
It deserves mention that non-minimally coupled scalar fields were utilised by 
Salopek et al\cite{salopek} and Spokoiny \cite{spokoiny1, spokoiny2} in the 
context of an early 
inflation, where the field had an oscillation or a plateau at the end of 
the inflation. In the present case, we need just the reverse for the Q-field, 
and that is aided by the non-minimally coupled field $\psi$.
\par We see that for a wide range of choice of $V(\phi)$, a power 
law acceleration is on cards, only the value of $V(\phi)$ at some 
initial stage is restricted by equation (16). For an exponential 
expansion of the scale factor, however, the potential $V(\phi)$ 
has to be an exponential function of $\phi$ ( in view of 
equation (22) ). This investigation can be extended for more 
complicated kinds of accelerated expansion. 
\par It is true that general 
relativity is by far the best theory of gravity and the present 
calculations are worked out in Brans-Dicke (B-D) theory, but this 
should give some idea about how a second scalar field may be 
conveniently used to get some desired results. Although B-D theory 
lost a part of its appeal as the most natural generalization of 
general relativity (GR) as the merger of B-D theory with GR for 
large $\omega$ limit is shown to be somewhat restricted 
\cite{ssen},
it still provides useful limits to the solution of the cosmological 
problems \cite{la}. Another feature of the present work is that the
numerical value of $\omega$ required is not much restricted. 
For power law expansion, $\omega$ is restricted by equation (16) 
which clearly shows that it can be adjusted by properly choosing 
values of some other quantities, whereas for exponential expansion,
equation (22) shows that $\omega$ is arbitrary. This is encouraging 
as it might be possible to get an acceleration even with a high value 
of $\omega$, compatible with local astronomical observations \cite{will}. 
B-D theory had been shown to generate acceleration by itself \cite{pavon}, 
although it had problems with early universe dynamics. B-D theory 
with quintessence or some modifications of the theory \cite{pav,sesh, aas, 
sen, ber} 
were shown to explain the present cosmic acceleration, but all 
these models, unlike the present work, required a very low value 
of $\omega$, contrary to the local observations.

\end{document}